\documentstyle[preprint,prl,aps]{revtex}
\begin{document}
\draft

\title{Hydrodynamic description of neutrino gas}

\author{Levan N. Tsintsadze}
\address{Venture Business Laboratory, Hiroshima University, Higashi-Hiroshima,
Japan}

\maketitle

\begin{abstract}
The system of neutrino-antineutrino $(\nu\bar{\nu})$ - plasma is
considered taking into account their weak Fermi interaction. New
fluid instabilities driven by strong neutrino flux in a plasma
are found. The nonlinear stationary as well as nonstationary waves
in the neutrino gas are discussed. It is shown that a bunch of
neutrinos, drifting with a constant velocity across a homogeneous
plasma, can also induce emission of lower energy neutrinos due to
scattering, i.e. the decay of a heavy neutrino $\nu_{H}$ into a
heavy and a light neutrino $\nu_L$ ($\nu_H\rightarrow\nu_H\nu_L$)
in a plasma. Furthermore we find that the neutrino production in
stars does not lead in general to energy losses from the neutron
stars.
\end{abstract}

\vspace{5cm}


\section{Introduction}
Our understanding of the properties of neutrinos in a plasma has
recently undergone some appreciable theoretical progress
\cite{Zel83,Ora,Tsin98,TsinP98,Vol}. The interaction of neutrinos
with a plasma particles, the creation of $\nu\bar{\nu}$ pairs, the
emission of neutrinos due to the collapse of a star are of primary
interests \cite{Zel83,Wil,Bet,Raf,Taj} in the description of some
astrophysical events such as supernova explosion. One of the key
processes upon the explosions are the large-scale hydrodynamic
instabilities as well as $\nu$ driven plasma instabilities. These
processes are also believed to have occurred during the lepton
stage of the early universe. During the formation of a neutron
star the collapsed core of the supernova is so dense and hot that
$\nu$ and $\bar{\nu}$ are trapped and are thus unable to leave the
core region of the neutron star. The rates of escape of $\nu$ and
$\bar{\nu}$ are very small, inside the star an equilibrium state
is reached, which includes the $\nu\bar{\nu}$ concentration.
Recently, the neutrino transport phenomenon during the
Kelvin-Helmholtz phase of birth of a neutron star in the diffusion
approximation was investigated by Pons et al. \cite{Pons}. A
detailed knowledge of transport properties of the neutrinos in
these extreme environments must include the neutrino-electron
Fermi weak interaction coupling. There are different mechanisms
involved in the creation of the $\nu$, as well as pairs. The
collective effects of the stellar plasmas can significantly alter
the production rate of neutrinos. It is widely thought that the
most dramatic plasma process is the decay of photons as plasmons
into neutrino pairs. It was pointed out by Adams et al. \cite{Ada}
that the neutrino pair radiation can be the dominant energy loss
mechanism for the plasma of the very dense stars, as well as white
dwarfs, red giants and supernovas. It has been also shown
\cite{Gvo} that a $\nu_H$ can undergo radiative decay into a
photon and a $\nu_L$ ($\nu_H\rightarrow\gamma\nu_L$) in the
presence of a strong magnetic field, with the strength greater
than the critical value given by $H_{cr}=m_e^2c^3/e\hbar=
4.41\cdot 10^{13}$ Gauss. The increase (decrease) of the $\nu$
($\bar{\nu}$) energy within the plasma was also demonstrated
\cite{TsinP98}. This may have a significant and potentially
detectable effect. Observationally, the recent results from the
Kamiokande group provide strong evidence for the existence of
neutrino oscillations. These results, together with the increased
confidence in ability to produce and manipulate intense muon beams
makes feasible the future neutrino factories based on muon storage
rings. A muon storage ring, as presently envisioned, would have
energy in the 10-50 GeV range, and produce directed beams of
intense neutrinos, and naturally such an intense stream of
neutrinos can exist in astrophysical and cosmological plasmas.
Thus, it is of interest to examine if new physical processes that
are caused by the intense flux of neutrinos, can appear in such a
plasma, i.e., of particular conceptual interest are effects which
have no counterpart in vacuum.

In this paper we consider a medium of neutrino gas and plasma in
semitransparent regions, where collective process plays an
important role. In our consideration we assume that the
interaction between neutrinos is weak in comparison with the
neutrino-electron interaction, i.e., the neutrino gas is ideal. We
can also ignore the spin of neutrinos in an isotropic plasma,
since in this case the energy of neutrinos is independent of the
spin operator. With this modeling, here we show that there exists
the opposite effect that has been proposed by others
\cite{Ora,Ada}: namely that the neutrinos production does not
generally lead to energy losses from a hot and dense system.

\section{Basic Equations}
In the following we shall demonstrate the excitation of
longitudinal oscillations in neutrino flux-plasma medium. For this
purpose, we make use of hydrodynamic equations for the neutrino
gas, which have been derived by Tsintsadze et al. \cite{TsinP98}.
These equations for the neutrino density $n_\nu$ and the fluid
velocity $\vec{u}_\nu$ take the following forms
\begin{eqnarray}
\label{cont n} \frac{\partial n_\nu}{\partial t}+div(n_\nu
\vec{u}_\nu) =0 \\ \label{momen} \frac{\partial
\vec{u}_\nu}{\partial t}+
(\vec{u}_\nu\cdot\vec{\nabla})\vec{u}_\nu= -\frac{Uc^2}{\hbar
<\omega_k>}\vec{\nabla} \frac{n_e}{n_{0e}},
\end{eqnarray}
where $U=\sqrt{2}G_Fn_{0e}$ describes the universal Fermi weak
interaction, $G_F=10^{-49}erg.cm^3$ is the Fermi coupling
constant,
$\frac{n_\nu}{<\omega_k>}=\int\frac{d^3k}{(2\pi)^3}\frac{N_k}{\omega_k}$,
and $n_e$ is the electron plasma density.

The electron continuity equation takes the form:
\begin{equation}
\label{cont e}
\frac{\partial n_e}{\partial t} +div(n_e\vec{v}_e)=0.
\end{equation}
For the equation of motion of electrons in the
presence of a neutrino gas, we have
\begin{equation}
\label{motion e} \frac{d\vec{p}_e}{dt}=e\nabla\phi
-\sqrt{2}G_F\nabla n_\nu \ ,
\end{equation}
where we have neglected the pressure term in comparison with the
first term on the right hand side under assumption that $r_{De}^2/
L^2\ll 1$, where $r_{De}$ is the Debye length of the electrons,
and L is the characteristic length of the system. Note that the
last term in Eq.(\ref{motion e}) is the neutrino ponderomotive
force. A more elaborate discussion on the neutrino ponderomotive
force is given in \cite{TsinP98}.

The electrostatic potential $\phi$ is determined from Poisson
equation
\begin{equation}
\label{poisson} \nabla^2\phi=4\pi e (n_e-n_{i})\ .
\end{equation}
We first consider a situation in which the ions are in an
equilibrium state, i.e. $n_i=n_{0i}$, where $n_{0i}$ is the
equilibrium value of the ion density. In this consideration the
range of frequency of small perturbation of quantities of the
medium is much less than the Langmuir frequency of the electrons.
Thus, we can neglect the inertial term in Eq.(\ref{motion e}), and
define $\phi$ from it. Finally, we obtain from the Poisson
equation the expression for the electron density
\begin{equation}
\label{eldens}
\frac{n_e}{n_{0e}}=1+\frac{\sqrt{2}G_{F}n_{0\nu}}{4\pi
e^2n_{0e}}\nabla^2\frac{n_\nu}{n_{0\nu}} \ ,
\end{equation}
where $n_{0\nu}$ is the equilibrium value of the neutrino density.
Substituting expression (\ref{eldens}) into Eq.(\ref{momen}) we
get the following equation, which except quantities of the
neutrinos contains only the electron charge
\begin{equation}
\label{mom1} \frac{\partial \vec{u}_\nu}{\partial t}+
(\vec{u}_\nu\cdot\vec{\nabla})\vec{u}_\nu=-\beta\vec{\nabla}(\nabla^2n_\nu)
\ ,
\end{equation}
where $\beta=G_{F}^2c^2n_{0\nu}/2\pi e^2\hbar <\omega_k>$.

Equations (\ref{cont n}) and (\ref{mom1}) form a closed set of
equations that describes the linear as well as nonlinear waves in
the neutrino gas.

\section{Cherenkov type of emission of light neutrinos}

Let us consider the propagation of small perturbations in a
homogeneous plasma-neutrino beam. To this end, we linearize Eqs.(\ref{cont n}) and
(\ref{mom1})
with respect to the perturbations, which are represented as
$ n_\nu=n_{0\nu}+\delta n_\nu,\
\vec{u}_{\nu}=\vec{u}_{0\nu}+\delta\vec{u}_\nu,\
$ where the suffix $0$ denotes the constant equilibrium value, and $\delta
n_\nu,\delta\vec{u}_\nu$ are small variations in the wave.
After linearization
of Eqs.(\ref{cont n}) and (\ref{mom1}), we will seek plane wave solutions proportional to
$exp i(\vec{q}\vec{r}
-\Omega t)$. We can then derive the following dispersion relation
\begin{equation}
\label{dis1}
(\Omega-\vec{q}\vec{u}_{0\nu})^2=-\beta q^4 \ .
\end{equation}
This leads to a solution of the form $\Omega=Re\Omega+iIm\Omega$ such that:
\begin{equation}
\label{re}
Re\Omega=qu_{0\nu}cos\Theta
\end{equation}
and the growth rate for the unstable branch becomes:
\begin{equation}
\label{im}
Im\Omega=\sqrt{\beta}q^2 \ .
\end{equation}
This solution clearly describes the emission of the low-frequency neutrinos
inside
a resonance cone ($\cos\theta=\frac{Re\Omega}{qu_{o\nu}}$), similar to
the well-known
Cherenkov emission of electromagnetic waves by charged particles moving in
uniform
medium with a velocity larger than the phase velocities of the emitted
waves.
On the other hand, strictly speaking the effect considered here is
physically quite
different from the usual Cherenkov emission by charged particles. The point
is that
the neutrino flux can no longer excite Langmuir waves and energetic (heavy)
neutrino scatters of the electron density perturbation (caused by the
ponderomotive force of neutrinos) and emits low (light) energy neutrino.
We specifically note here that such decay process $\nu_H\rightarrow\nu_H\nu_L$
is forbidden without a plasma.

Let us estimate the growth rate (\ref{im}) for some typical
neutrino parameters. Accordingly, we take the $\nu_L$ energy to
be about 100 keV, the $\nu_H$ energy to be about 1-10 MeV, and
$n_\nu\sim 10^{32}$. It is important to emphasize that the growth
rate does not depend on the density and temperature of a plasma.
Then for the growth rate (\ref{im}) we obtain $Im\Omega\simeq
1.7-5.4\cdot 10^8 sec^{-1}$. Comparing this with the collision
frequency $\delta$ of neutrinos, it can be seen the latter to be
much less than the former. Indeed, for the given mean free path
$l_\nu=2km(10/\varepsilon_\nu)^2=2\cdot 10^5$cm \cite{Bet}, the
collision frequency is
$\delta\sim\frac{<v>}{l_\nu}\sim\frac{c}{l_\nu}\sim 10^5
sec^{-1}$, i.e. under these circumstances, the medium for the
neutrino is "collisionless".

\section{Nonlinear waves}

We now discuss the nonlinear features of evolution of the
unstable waves. If we consider the case, when the period of
oscillations, $1/Re\Omega$, is much less than the wave amplitude
growth time, $1/Im\Omega$, then the dispersion relation can be
written in a one-dimensional perturbation as $\Omega=qu_{0\nu}$.
This corresponds to the condition for the neglect of the third
derivatives in Eq.(\ref{mom1}). Doing so, we obtain
\begin{equation}
\label{mom2} \frac{\partial u_\nu}{\partial
t}+u_\nu\frac{\partial}{\partial z}u_\nu =0 \ ,
\end{equation}
which describes a beam of noninteracting neutrinos.

As was shown in \cite{Kad}, the equation type of Eq.(\ref{mom2})
has solutions with the characteristic properties of nonlinear
waves. It has been verified that the initial perturbation of the
velocity $(u(z,0)\sim sin kz)$ in the phase plane $(u,z)$ after
some time leads the wave $u(z,t)$ to break.

Thus, our results indicate that the neutrino beam of
noninteracting particles, similar to a beam of plasma particles,
has many properties of nonlinear system. Namely, it is subject to
wave breaking and to generation of higher harmonics.

If we take into account the temperature $T_\nu$ of neutrinos and suppose that
the neutrino gas is an ideal ($P_\nu=n_\nu
T_\nu=P_{0\nu}(n_\nu/n_{0\nu})^\gamma$), then the equation of motion
(\ref{mom1})
modifies as
\begin{equation}
\label{momT}
\frac{d\vec{u}_\nu}{dt}=-\frac{C_\nu^2(n_\nu)}{n_\nu}\vec{\nabla}n_\nu-
\beta\vec{\nabla}(\nabla^2n_\nu) \ ,
\end{equation}
where $C_\nu^2(n_\nu)=C_{0\nu}^2(n_\nu/n_{0\nu})^{\gamma-1}$, $C_{0\nu}$ is
the velocity of sound in the neutrino gas, and $\gamma$ is the effective
adiabatic index.

Equations (\ref{cont n}) and (\ref{momT}) are so called Boussinesq
equations. We can see that the character of nonlinear processes
strongly depends on the dispersion, i.e., the dependence of phase
velocity on the wave number. The most important nonlinear effect
here, as it was shown above, is that the wave front steepens. At
this point the dispersion becomes significant. Waves having
different wave number have different phase velocities, and the
nonlinear steepening of front can be compensated by the dispersive
spread. Therefore, in this case the stationary waves can exist,
which propagate with constant velocities without changing their
shape (e.g., solitons). To note, Eqs.(\ref{cont n}) and
(\ref{momT}) describe also the nonstationary waves \cite{Gar}.

Equations type of Eqs.(\ref{cont n}) and (\ref{momT}) have been thoroughly
studied, and it has been shown that these kind of equations have many periodic
as well as solitary solutions. If we use the common method for the weak
dispersive medium, we obtain from Eqs.(\ref{cont n}) and (\ref{momT}) the
Korteweg-de Vries equation for the neutrino velocity
\begin{equation}
\label{kor} \frac{\partial u}{\partial
t}+u\frac{\partial}{\partial z}u+\alpha\frac{\partial^3}{\partial
z^3}u=0 \ ,
\end{equation}
where $u=\frac{(\gamma +1)}{2}u_\nu$, and $\alpha=\frac{\beta}{2C_{0\nu}}$.
This equation is well studied and we do not repeat all solutions and
conservation laws of it. Our aim was to prove that the neutrino gas exhibits
the same features as the waves in plasmas (e.g. electromagnetic waves, waves
on the surface of a heavy liquid, waves in biology, etc.).

\section{Ion dynamics and instabilities}

We next consider the excitation of an ion-sound waves by the neutrino flux.
Above we studied the high-frequency oscillations neglecting the ion dynamics.
However, the ion impact turns out to be extremely important in the
low-frequency waves which may be excited in a strong non-isothermal plasma with
hot electrons and cold ions. As for the electrons we may assume that the
electrons are in equilibrium under the conditions of low-frequency oscillation
and that their density is determined by the Boltzmann formula:
\begin{equation}
\label{bol}
n_e=n_{0e}e^{\frac{e\phi-\sqrt{2}G_F\delta n_\nu}{T_e}}\ , \hspace{2cm}
\delta n_\nu=n_\nu-n_{0\nu} \ .
\end{equation}
The ions can be described hydrodynamically by means of the HD equations
($T_i=0$):
\begin{equation}
\label{ion}
m_i\frac{dv_i}{dt}=-e\nabla\phi\ , \hspace{2cm} \frac{\partial n_i}{\partial
t}+div (n_i\vec{v}_i)=0 \ ,
\end{equation}
and the Poisson equation in this case is
\begin{equation}
\label{pois2}
\nabla^2\phi =4\pi e(n_{0e}e^{\frac{e\phi-\sqrt{2}G_F\delta n_\nu}{T_e}}-n_i) \
.
\end{equation}
After linearization of the set of Eqs.(\ref{cont
n}),(\ref{momen}),(\ref{ion}) and (\ref{pois2}) we obtain the
dispersion relation for the ion-sound oscillation in the presence
of the monoenergetic neutrino beam as
\begin{equation}
\label{disI}
(\Omega^2-\omega_s^2)\Bigl\{(\Omega-\vec{q}\vec{u}_{0\nu})^2+\frac{q^2r_D^2}{1+q^2r_D^2}
\alpha q^2c^2\Bigr\}=\frac{\omega_s^2\alpha q^2c^2}{1+q^2r_D^2} \ ,
\end{equation}
where $\alpha=\frac{2G_F^2n_{0e}n_{0\nu}}{\hbar <\omega_k>T_e}$,
and  $\omega_s$ is simply the ion-sound frequency,
\begin{equation}
\label{oms}
\omega_s=\frac{qv_s}{\sqrt{1+q^2r_D^2}} \ ,
\end{equation}
here $v_s=\sqrt{\frac{T_e}{m_i}}$ is the ion-sound velocity.

We examine the dispersion relation (\ref{disI}) for two interesting cases.
Let us first discuss the small wavelength case, $qr_D\gg 1$ and
$\Omega\ll\omega_s$. In this case Eq.(\ref{disI}) reduces
to
\begin{equation}
\label{disI1}
(\Omega-\vec{q}\vec{u}_{0\nu})^2+\alpha q^2c^2=0 \ ,
\end{equation}
which has a solution of the form $\Omega=Re\Omega+iIm\Omega$ with
\begin{equation}
\label{re1}
Re\Omega=qu_{0\nu}cos\theta \hspace{2cm} and \hspace{2cm}
Im\Omega=\sqrt{\alpha}qc \ .
\end{equation}
Hence, this solution describes the emission of low-frequency
neutrinos inside a resonance cone ($cos\theta$).

Next, for the long wavelengths, $qr_D\ll 1$, the dispersion relation
(\ref{disI}) casts in to the form
\begin{equation}
\label{disI2}
(\Omega^2-q^2v_s^2)(\Omega-\vec{q}\vec{u}_{0\nu})^2-\alpha
q^2v_s^2q^2c^2=0 \ .
\end{equation}
Here we look for the solution with coincide roots, i.e. $\Omega=qv_s+\Gamma$
and $\Omega=\vec{q}\vec{u}_{0\nu}+\Gamma$, and get for the growth rate of the
oscillatory instabilities the following expression
\begin{equation}
\label{im2}
Im\Gamma=\frac{\sqrt{3}}{2}\Bigl(\frac{v_s}{2c}\alpha\Bigr)^{1/3}qc \ .
\end{equation}

\section{Formation of shock waves}

Let us now investigate the nonlinear low-frequency waves. To this
end, we add the ion pressure to the equation of motion of the
ions, Eq.(\ref{ion}), and since $\Omega\ll\omega_s$ neglect the
inertial term. Then solution of this equation is
\begin{equation}
\label{dens}
 n_i=n_{0i}e^{-\frac{e\phi}{T_i}} \ .
\end{equation}
If we further impose the quasi-neutrality, $n_e\simeq n_i$, we obtain
\begin{equation}
\label{phi}
e\phi=\sqrt{2}G_F\delta n_\nu\frac{T_i}{T_e+T_i} \ .
\end{equation}
Substituting (\ref{phi}) into the Boltzmann distribution function
of the electrons (\ref{bol}), we get
\begin{equation}
\label{den e}
\frac{n_e}{n_{0e}}=e^{-\frac{\sqrt{2}G_F\delta n_\nu}{T_e+T_i}}\ .
\end{equation}
For the neutrino gas the natural requirement is $G_F\delta n_\nu\ll T_e$, so
that we have
\begin{equation}
\label{den e1}
\frac{n_e-n_{0e}}{n_{0e}}=-\frac{\sqrt{2}G_F\delta n_\nu}{T_e+T_i}\ .
\end{equation}
This equation when substituted into Eq.(\ref{momen}) yields the equation of
motion of the neutrinos
\begin{equation}
\label{mom1n} \frac{\partial \vec{u}_\nu}{\partial t}+
(\vec{u}_\nu\cdot\vec{\nabla})\vec{u}_\nu=\delta\vec{\nabla}\frac{n_\nu}{n_{0\nu}}
\ ,
\end{equation}
where $\delta=\frac{2G_F^2c^2n_{0\nu}n_{0e}}{\hbar <\omega_k>(T_e+T_i)}$. If we
linearize this equation and the equation of density (\ref{cont n}), then we
obtain the dispersion equation (\ref{disI1}) for $T_i=0$.

We now consider the one-dimensional nonlinear waves, for this case equations
are
\begin{eqnarray}
\label{1dcon}
\frac{\partial n_\nu}{\partial t}+\frac{\partial}{\partial z}n_\nu u_\nu=0 \\
\label{1dmom} \frac{\partial u_\nu}{\partial
t}+u_\nu\frac{\partial}{\partial
z}u_\nu=\delta\frac{\partial}{\partial z}\frac{n_\nu}{n_{0\nu}} \
.
\end{eqnarray}
These equations have exact solutions for the initial profile of
the neutrino density
\begin{equation}
\label{nden} n_\nu(z,0)=\frac{n_{0\nu}}{ch^2\frac{z}{z_0}}\ ,
\end{equation}
and in nondimensional quantities are written as
\begin{eqnarray}
\label{ro} Q=(1+Q^2\tau^2)ch^{-2}(\xi-V\tau) \\ \label{V}
V=-2Q\tau th(\xi-V\tau) \ ,
\end{eqnarray}
where $Q=\frac{n_\nu(z,t)}{n_{0\nu}}$, $\tau=\omega_{pi} t$,
$\xi=\frac{\omega_{pi}}{\sqrt{\delta}}z$,
$V=\frac{u_\nu}{\sqrt{\delta}}$, with
$\omega_{pi}=\Bigl(\frac{4\pi e^2n_{0e}}{m_i}\Bigr)^{1/2}$ being
the Langmuir frequency of ions.

From Eqs.(\ref{ro}) and (\ref{V}), we can readily derive
\begin{equation}
\label{xi}
\xi=arch\sqrt{\frac{1+Q^2\tau^2}{Q}}-2Q\tau^2\sqrt{1-\frac{Q}{1+Q^2\tau^2}}
\ .
\end{equation}
Examining Eq.(\ref{xi}), we now  discuss the question how the
maximum of the initial density changes in time. For this purpose,
we take a derivative of Eq.(\ref{xi}) by $Q$, i.e.
$\frac{d\xi}{dQ}\mid_{\xi=0}$, and allow it to be infinite. From
this condition, we find the expression for the maximum density
\begin{equation}
\label{max} Q_{max}(0,\tau)=\frac{1}{2\tau^2}(1-\sqrt{1-4\tau^2})
\ .
\end{equation}
One can see that at $\tau\rightarrow 0$, $Q_{max}\rightarrow 1$,
$n_\nu(0,0)=n_{0\nu}$. Then with increase of time ($\tau$) the
shape of the density changes, namely becomes narrow. Also, the
center of the density ($\xi=0$) increases with rise of $\tau$ and
reaches another maximum, $Q_{max}=2$ at $\tau=1/2$. After a
certain value $\tau=\tau_0$, the shock waves are formed, and
derivatives are $ \frac{\partial
Q}{\partial\xi}\mid_{\xi=0,\tau=\tau_0}=\infty \ ,$ \hspace{.2cm}
$\frac{\partial^2 Q}{\partial\xi^2}\mid_{\xi=0,\tau=\tau_0}=\infty
. \ $ However, the density itself remains continuous. Using these
conditions from Eq.(\ref{xi}) follows the relation
$Q_0\tau_0^2=3/2$, where $\tau_0>0.5$ and $Q_0^2\tau_0^2\gg 1$.

3D numerical simulation which underscores the above picture is
carried out. Namely, we consider 3D problem, when the initial
profile of density of neutrinos is the Gaussian distribution
\begin{equation}
\label{gauss}
n_\nu(\vec{r},0)=n_{0\nu}exp\Bigl\{-\frac{x^2+y^2}{2a_0^2}-\frac{z^2}{2z_0^2}\Bigr\}
\ .
\end{equation}
Numerical studies have shown that the dynamics of non-linearity is
determined by the dimensionality of problem and of the initial
profile of the neutrino density.

In order to confirm the validity of 1D analytical calculation, we
first discuss 3D numerical analysis of Eqs.(\ref{cont n}) and
(\ref{mom1n}) for pancake-shaped beams, i.e., $a_0=2$, and
$z_0=0.5$. Figures 1a and 1b show that the width of the density
becomes shorter in time along the propagation direction (z-axis).
The density of neutrinos is concentrated in the small region (on
the z-axis), and at $t=4$ starts to form the shock wave. Whereas
Figures 1c and 1d show the insignificant expansion of the pancake
beams in the transverse direction. Therefore, we can conclude that
the 1D analytical calculation is very good approximation for the
3D pancake-shaped beams.

Quite different processes develop when the initial distribution of
neutrinos has a form of the bullet, i.e., when $a_0=0.5$ and
$z=2$. We can see that there are two special competitive regimes:
one, which is due to the self-focusing, is the compression of the
neutrino density in the transverse direction (see Figures 2a and
2b), and the second is the expansion of the density along the
propagation direction (see Figures 2b and 2c). The redistribution
of neutrinos in the transverse direction is accompanied by the
formation of sharp maximum of density near the z-axis
($r=\sqrt{x^2+y^2}\rightarrow 0$), and as Figure 2a shows, the
formation of the shock wave takes place at t=2.8.

\section{SUMMARY AND DISCUSSIONS}

We have studied the problem of weak Fermi interaction of a
neutrino beam with a plasma. Novel fluid instabilities driven by
the neutrino flux in a plasma are observed. The range of
wavelengths corresponding to several instabilities, discussed in
this paper, for relevant parameters of the neutrino gas and plasma
is $\lambda\sim 1-10^{-9}cm$. We have found that for the neutrino
beam there exists a Cherenkov type of emission of low energy
neutrinos; this is different from the usual Cherenkov effect,
though. It should be emphasized that in our case the physical
mechanism is that the high energy neutrino does not just excite
the plasma wave but also can scatter from density perturbations of
electrons and can emit the low energy neutrino. We also found that
in the usual hot plasma environments such as supernova explosions,
the effect considered above prevents neutrino escape. Since the
low energy neutrinos cannot stream away from the central regions
of the plasma one can have an accumulation of such neutrinos in a
dense stellar medium. Thus, the above discussed processes can
considerably modify certain phases of evolution in a stellar
model. It should be emphasized that unlike the previous models
\cite{Ada},\cite{chi} on neutrino pair emission by a stellar
plasma, the emission of low energy neutrinos by high energy
neutrinos, as discussed in this paper, does not depend on the
density and temperature of a plasma. The growth rate found for
some typical neutrino parameters is $5.4\cdot 10^8 sec^{-1}$.
Which is much larger than the collision frequency of neutrinos,
$10^5 sec^{-1}$. Thus in this case the medium for the neutrino is
"collisionless".

The nonlinear stationary as well as nonstationary waves in the
neutrino gas are also discussed. The dispersive effect, which is
responsible for the existence of the nonlinear stationary waves,
e.g. solitons, in the neutrino gas is due to the electron density
modulation by the neutrino ponderomotive force \cite{TsinP98}. Our
results indicate that a neutrino beam is subject to wave breaking
and shock wave formation. Thus, high energy neutrinos will lose
energy also by wave breaking in addition to Cherenkov emission,
leading to plasma heating (and neutrino cooling). Whereas, the
shock waves can produce a relativistic energy flow. In addition,
the solitons can also be a potential candidates for the generation
of relativistic particles. These particles then release the energy
and produce the observed radiation in gamma-ray bursts (GRBs).
Hence, one has a origin of radiation in the modeling of burst
sources, which requires a discussion of particle acceleration
processes. Since the simplest, most conventional, and practically
inevitable interpretation of the observations of GRBs is that GRBs
result from the conversion of the kinetic energy of
ultra-relativistic particles to radiation. It is well accepted
that many of them originate in the very distant, early universe.
In the early universe, the processes discussed in the paper can
also lead to formation of nonlinear structures, contributing to
the formation of the large scale structure of the Universe
\cite{Mis}. Which, is believed, grew gravitationally out of small
density fluctuations \cite{wei}. The effect of this density
variation in the early universe was left on the cosmic microwave
background radiation in the form of spatial temperature
fluctuations. It should be emphasized that the gravity, however,
cannot produce these fluctuations, but increase alone. Therefore a
discussion of the physical models of generation of the initial
matter density fluctuations is very crucial. There is no
comprehensive model at present that can explain their origin. The
weak Fermi interaction in a plasma, as discussed in this paper,
can be an alternative and a new source for the required density
perturbations. As demonstrated, a long lived nonlinear structures,
which carry large amount of mass and energy, are generated in such
a system. Since an initial localization of mass and energy is
exactly that the gravity needs for eventual structure formation,
weak Fermi interaction may have provided a decisive element in the
formation of a large scale map of the observable Universe.

\appendix
\section{The generalization of the Wigner-Moyal equation}

The Wigner-Moyal equation in quantum kinetic theory is the
analogy of the one-particle Liouville equation in kinetic theory.
In order to derive this equation, we start with a dispersion
relation of a single neutrino, taking into account the universal
weak Fermi interaction between plasma electrons and neutrinos
($U=\sqrt{2}G_Fn_e$)
\begin{eqnarray}
(E-U)^{2}=p^{2}c^{2}+m_{\nu}^{2}c^{4} \ , \label{dispF}
\end{eqnarray}
where E is the energy, p is the momentum,  $m_{\nu}$ is the
neutrino rest mass. This equation has two distinct energy
solutions
\begin{eqnarray}
E=U\pm c\Bigl(p^{2}+m_{\nu}^{2}c^{2}\Bigr)^{1/2} \ . \label{ensol}
\end{eqnarray}
The physical meaning of this result is best understood when we
are in the rest frame of the background medium, which creates the
potential energy U. The positive sign in Eq.(\ref{ensol})
corresponds to the neutrino solution and it shows that, for a
given magnitude of the particle momentum, the energy of the
neutrino is increased by an amount U with respect to its value in
vacuum. The other solution in Eq.(\ref{ensol}), corresponding to
the negative sign, is that of an antineutrino solution.

As mentioned in the Introduction, in an isotropic plasma one can
ignore the spin of neutrinos, and therefore we can consider a
scalar wave function $\psi_{\nu}$, associated with the neutrino
gas, and we can write in equilibrium
\begin{eqnarray}
\Bigl\{\omega-c(k^{2}+k_{0}^{2})^{1/2}-\omega_{F}\Bigr\}\psi_{\nu}=0
\ . \label{sfan}
\end{eqnarray}
where $\omega=\frac{E}{\hbar}$, $\vec{k}=\frac{\vec{p}}{\hbar}$,
$k_0=\frac{m_\nu c}{\hbar}$ and $\omega_{F}=\frac{U}{\hbar}$ is
the Fermi frequency.

On the other hand, if the neutrino gas is not in equilibrium, the
amplitude of $\psi_{\nu} $ will slowly change in space and time
due to the interaction of neutrinos with the background plasma
fluctuations. We can then use the geometrical optic
approximation, $\omega\rightarrow\omega -i\partial/\partial t, $
$\vec{k}\rightarrow \vec{k}+i\vec{\nabla}$ and
$n_{e}=n_{0e}+\delta n_{e}$
($\omega\gg\mid\frac{\partial}{\partial t}\mid$,
$\mid\vec{k}\mid\gg\mid\nabla\mid$). Doing expansion
\begin{eqnarray}
\sqrt{(\vec{k}+i\vec{\nabla})^2+k_0^2}=\sum_{j=0}^{N}\Bigl(_j^N\Bigr)(k^2+k_0^2)^{N-j}
(2i\vec{k}\vec{\nabla}-\nabla^2)^j  \label{expen}
\end{eqnarray}
and neglecting higher order derivatives greater than $\nabla^{2}$
in Eq.(\ref{expen}), we arrive  from Eq.(\ref{sfan}) to
\begin{eqnarray}
i\Bigl(\frac{\partial}{\partial
t}+\vec{v}_{g}\vec{\nabla}\Bigr)\psi_{\nu}-
\frac{v_{g}}{2k}\Bigl(\nabla^{2}-\Bigl(\frac{\vec{v}_{g}}{c}\vec{\nabla}
\Bigr)^{2}\Bigr)\psi_{\nu}-\omega_{F}\frac{\delta
n_{e}}{n_{0e}}\psi_{\nu}=0 \ , \label{apkin}
\end{eqnarray}
where use was made of the relation
$\omega=c(k^{2}+k_{0}^{2})^{1/2}+\omega_{F}(n_{0e})$, and the
following notations were introduced
\begin{eqnarray}
\vec{v}_{g}=\frac{c\vec{k}}{(k^{2}+k_{0}^{2})^{1/2}}\ ,
\hspace{.5cm} \Bigl(\frac{\partial\omega_{F}}{\partial
n}\Bigr)_{n=n_{0e}}=\frac{\omega_{F}(n_ {0e})}{n_{0e}} \ ,
\end{eqnarray}
We now write the product of two wave functions $\psi_{\nu}$, which
are defined at two distinct points and instants of time
\begin{eqnarray}
F=\psi_{\nu}(\vec{r}_{1},t_{1})
\psi_{\nu}^{*}(\vec{r}_{2},t_{2})=\psi_{\nu}\Bigl(\vec{R}+\frac{\vec{r}}{2},
t+\frac{\tau}{2}\Bigr)\psi_{\nu}^{*}\Bigl(\vec{R}-\frac{\vec{r}}{2},
t-\frac{\tau}{2}\Bigr)\ , \label{twops}
\end{eqnarray}
where we have introduced new space and time variables defined by
\begin{eqnarray*}
\vec{R}=\frac{1}{2}(\vec{r}_{1}+\vec{r}_{2}), \hspace{.3cm}
\vec{r}=\vec{r}_{1}-\vec{r}_{2}, \hspace{.3cm}
t=\frac{1}{2}(t_{1}+t_{2}), \hspace{.3cm} \tau=t_{1}-t_{2} \ .
\end{eqnarray*}
Let us make a Fourier transform of (\ref{twops}) on the variables
$\vec{r}, \tau$ in order to introduce the Wigner function
$f(\vec{R},t,\vec{k},\omega)$
\begin{eqnarray}
f(\vec{R},t,\vec{k},\omega)=\frac{1}{(2\pi)^4}\int d\vec{r}\int
d\tau \psi_{\nu}\Bigl(\vec{R}+\frac{\vec{r}}{2},
t+\frac{\tau}{2}\Bigr)\psi_{\nu}^{*}\Bigl(\vec{R}-\frac{\vec{r}}{2},
t-\frac{\tau}{2}\Bigr) e^{-i(\vec{k}\vec{r}- \omega\tau)} \ .
\label{wiggen}
\end{eqnarray}
This is the generalized Wigner distribution function. Integration
of (\ref{wiggen}) by the frequency $\omega$ leads to the ordinary
expression  of the Wigner distribution function \cite{wig} as
\begin{eqnarray}
f^w(\vec{R},t,\vec{k})=\frac{1}{(2\pi)^3}\int d\vec{r}
\psi_{\nu}\Bigl(\vec{R}+\frac{\vec{r}}{2},
t\Bigr)\psi_{\nu}^{*}\Bigl(\vec{R}-\frac{\vec{r}}{2}, t\Bigr)
e^{-i\vec{k}\vec{r}} \ . \label{wigor}
\end{eqnarray}
From (\ref{wiggen}) also follows important relation
\begin{eqnarray}
\mid\psi_{\nu}\mid^2=\int\frac{d\vec{k}}{(2\pi)^3}\int\frac{d\omega}{2\pi}
f(\vec{R},t,\vec{k},\omega)\ .
\end{eqnarray}
Assuming that the neutrinos verify the dispersion relation
(\ref{ensol}), we can say that for a given value $\vec{k}$ we
always have a well-defined value of the frequency,
$\omega=\omega(k)$. This means that we can write
\begin{eqnarray}
f(\vec{R},t,\vec{k},\omega)=2\pi
f(\vec{R},t,\vec{k})\delta(\omega-\omega(k))\ .
\end{eqnarray}
Integration over the wave vector spectrum leads to the density of
neutrinos
\begin{eqnarray}
n_{\nu}(\vec{R},t)=\mid\psi_{\nu}\mid^2=\int\frac{d\vec{k}}{(2\pi
)^{3}}f(\vec{R},t,\vec{k}) \ .
\end{eqnarray}
Following the procedure described in previous works
\cite{wig,moy,tap,ltsin,TsinP98}, we can derive an equation for
the function (\ref{twops}), the result is
\begin{eqnarray}
i\Bigl\{\frac{\partial}{\partial
t}+\vec{v}_{g}\vec{\nabla}_{R}\Bigr\}F
-\frac{v_{g}}{2k}\Bigl\{\vec{\nabla}_{R}\cdot\vec{\nabla}_{r}-
\Bigl(\frac{\vec{v}_{g}}{c}\vec{\nabla}_{R}\Bigr)\Bigl(\frac{\vec{v}_{g}}{c}
\vec{\nabla}_{r}\Bigr)\Bigr\}F-\frac{\omega_{F}}{n_{0e}}(\delta
n_{1}-\delta n_{2})F=0 \ . \label{func}
\end{eqnarray}
Now making the Fourier transformation of Eq.(\ref{func}), we
obtain an evolution equation for function (\ref{wiggen}) in the
form
\begin{eqnarray}
\Bigl(\frac{\partial}{\partial
t}+\vec{v}_{g}^{eff}\vec{\nabla}_{R}\Bigr)f(\vec{R},t,\vec{k},\omega)-\frac{\omega_{F}}{n_{0e}}
\int d\vec{r}\int d\tau(\delta n_{1}-\delta n_{2})F
e^{-i(\vec{k}\vec{r}- \omega\tau)}=0 \ , \label{evol}
\end{eqnarray}
where
\begin{eqnarray*}
\vec{v}_{g}^{eff}=\vec{v}_{g}\Bigl(1+\frac{k_{0}^{2}}{
k^{2}+k_{0}^{2}}\Bigr) \ .
\end{eqnarray*}
If all derivatives of $\delta n$ exist, we can expand $\delta
n_{1}(\vec{R}+\frac{\vec{r}}{2}, t+\frac{\tau}{2})-\delta
n_{2}(\vec{R}-\frac{\vec{r}}{2}, t-\frac{\tau}{2})$ around
$(\vec{R},t)$, i.e.
\begin{eqnarray}
\delta n_{1}-\delta
n_{2}=\sum_{l=0}^{\infty}\frac{1}{l!}\Bigl(\frac{\vec{r}}{2}\cdot
\vec{\nabla}_R+\tau\frac{\partial}{\partial
t}\Bigr)^l[1-(-1)^l]\delta n(\vec{R},t)
\end{eqnarray}
and obtain the generalized Wigner-Moyal equation
\begin{eqnarray}
\Bigl(\frac{\partial}{\partial
t}+\vec{v}_{g}^{eff}\vec{\nabla}_{R}\Bigr)f(\vec{R},t,\vec{k},\omega)=\frac{\omega_{F}}{n_{0e}}
sin\Bigl\{\vec{\nabla}_{R}\cdot\vec{\nabla}_{k}-\frac{\partial
}{\partial t}\cdot\frac{\partial}{\partial\omega}\Bigr)\cdot\delta
n(\vec{R},t)f(\vec{R},t,\vec{k},\omega)\ . \label{moyalg}
\end{eqnarray}
Integration of Eq.(\ref{moyalg}) by the frequency leads to the
Wigner-Moyal equation for the Wigner function (\ref{wigor})
without the last term on the right hand side of Eq.(\ref{moyalg}).

If we just retain the first term on the right hand side of
Eq.(\ref{moyalg}), i.e., the term corresponding to $sinx\approx
x$, we obtain the well-known Liouville-Vlasov kinetic equation
for the distribution function (\ref{wiggen}), but with an
additional time derivative term
\begin{eqnarray}
\Bigl(\frac{\partial}{\partial
t}+\vec{v}_{g}^{eff}\vec{\nabla}_{R}\Bigr)f(\vec{R},t,\vec{k},\omega)-\frac{\omega_{F}}{n_{0e}}
\Bigl\{\vec{\nabla}_{R}\delta
n\cdot\vec{\nabla}_{k}f-\frac{\partial\delta n }{\partial
t}\cdot\frac{\partial f}{\partial\omega}\Bigr)=0 \ . \label{vlas}
\end{eqnarray}
The stationary solution of Eq.(\ref{vlas}) is the Fermi
distribution function
\begin{eqnarray}
f_F=\Bigl(e^{\frac{\hbar
c\sqrt{k^2+k_0^2}+U}{T_\nu}}+1\Bigr)^{-1}\ .
\end{eqnarray}
From Eq.(\ref{vlas}) we can now derive a set of fluid equations
for the neutrino gas. To this end we shall introduce the
principal definitions of quantities which describe the state of
the neutrinos in a plasma. Namely, the total number of neutrinos
is defined as
\begin{eqnarray}
N_\nu=\int d\vec{R}
\int\frac{d\vec{k}}{(2\pi)^3}\int\frac{d\omega}{2\pi}
f(\vec{R},t,\vec{k},\omega)=\int d\vec{R}n_\nu(\vec{R},t)\ ,
\label{numb}
\end{eqnarray}
the total energy of the neutrino gas
\begin{eqnarray}
E_{total}=\int d\vec{R}
\int\frac{d\vec{k}}{(2\pi)^3}\int\frac{d\omega}{2\pi} \hbar\omega
f(\vec{R},t,\vec{k},\omega)=\int d\vec{R}\varepsilon(\vec{R},t)\ ,
\label{total}
\end{eqnarray}
and the neutrino mean velocity
\begin{eqnarray}
\vec{u}_{\nu}=\frac{1}{n_{\nu}}\int\frac{d\vec{k}}{(2\pi)^{3}}
\int\frac{d\omega}{2\pi}
\frac{\vec{k}c^{2}}{\omega}f(\vec{R},t,\vec{k},\omega) \ .
\label{mean}
\end{eqnarray}
Having definitions (\ref{numb}-\ref{mean}) and using the usual
momentum procedure, from Eq.(\ref{vlas}) we obtain a set of fluid
equations
\begin{eqnarray}
\frac{\partial n_{\nu}}{\partial t}+div (n_{\nu}\vec{u}_{\nu})=0\
, \label{apcon}
\end{eqnarray}
\begin{eqnarray}
\frac{\partial \vec{u}_{\nu}}{\partial
t}+(\vec{u}_{\nu}\cdot\vec{\nabla})\vec{
u}_{\nu}=-\frac{\omega_{F}c^{2}}{<\omega_k>}\Bigl(\vec{\nabla}+
\frac{\vec{u}_{\nu}}{c^{2}}\frac{\partial}{\partial
t}\Bigr)\frac{\delta n_e}{n_{0e}}- \frac{ \nabla P_\nu}{n_{\nu}} \
, \label{apmom}
\end{eqnarray}
\begin{eqnarray}
\frac{\partial \varepsilon_\nu}{\partial t}+\vec{\nabla}\cdot
\vec{p}_\nu=\hbar\omega_{F}n_{\nu }\frac{\partial}{\partial
t}\frac{\delta n_e}{n_{0e}} \ , \label{apen}
\end{eqnarray}
where we used the definitions of the momentum density
\begin{eqnarray}
\vec{p}_\nu=\hbar\int\frac{d\vec{k}}{(2\pi)^{3}}\int\frac{d\omega}{2\pi}
\vec{k}f(\vec{R},t,\vec{k},\omega) \ ,
\end{eqnarray}
and
\begin{eqnarray}
\frac{n_\nu}{<\omega_k>}=\int\frac{d\vec{k}}{(2\pi)^{3}}\int\frac{d\omega}{2\pi}
\frac{f(\vec{R},t,\vec{k},\omega)}{\omega} \ .
\end{eqnarray}
The quantity $P_\nu$ plays  role of the neutrino pressure and is
defined as
\begin{eqnarray}
P_\nu=\frac{1}{3}\int\frac{d\vec{k}}{(2\pi)^{3}}\int\frac{d\omega}{2\pi}
\Bigl(\frac{\vec{k}c^{2}}{\omega}-\vec{u}_{\nu}\Bigr)^{2}f(\vec{R},t,\vec{k},\omega)
\ .
\end{eqnarray}
Equations (\ref{apcon}-\ref{apen}) together with equations of the
plasma, given in this paper, describe the collective process,
which plays a more important role than the pair collision effect.


\begin{figure}
\caption{a) and b) The time evolution of the neutrino density
along the z - axis in the case of the pancake-shaped beams\\
c) and d) The time evolution of the neutrino density along the x -
axis in the case of the pancake-shaped beams \label{fig1}}
\end{figure}

\begin{figure}
\caption{a) The time evolution of the neutrino density along the x
- axis in the case of the neutrino bullet \\
b) The neutrino density at t=2.8 in (z,x) space\\
c) The time evolution of the neutrino density along the z - axis
in the case of the neutrino bullet \label{fig2}}
\end{figure}


\begin{references}
\bibitem[*]{byline} Permanent address: Institute of Physics,
Georgian Academy of Sciences, Tbilisi 77.

\bibitem{Zel83}
Ya.B.Zel'dovich, and I.D.Novikov, {\sl Relativistic Astrophysics}
(University of Chicago Press, Chicago, 1983).

\bibitem{Ora}
V.N.Oraevski, and V.S.Semikoz, Sov. Phys. JETP {\bf 59}, 465
(1984); V.N.Oraevski, et al., ibid, {\bf 66}, 890 (1987);
J.C.D'Olivo, and P.B.Pal, Phys. Rev. D {\bf 49}, 1398 (1994).

\bibitem{Tsin98}
N.L.Tsintsadze, J.T.Mendonca, P.K.Shukla, Phys. Lett. A {\bf 249},
110 (1998); J.T.Mendonca, et al., ibid, {\bf 239}, 373 (1998);
N.L.Tsintsadze,  P.K.Shukla,  J.T.Mendonca, and L.N.Tsintsadze,
Phys. Scripta {\bf T82}, 128 (1999).

\bibitem{TsinP98}
N.L.Tsintsadze, J.T.Mendonca, and L.N.Tsintsadze, Phys. Plasmas
{\bf 5}, 3512 (1998).

\bibitem{Vol}
M.B.Voloshin, et al., Sov. Phys. JETP {\bf 64}, 446 (1986);
I.Daicic, et al., Phys. Rep. {\bf 237}, 65 (1994); Review of
Particle Physics, par.1 Phys. Rev. D {\bf 54}, 282 (1996).

\bibitem{Wil}
J.R.Wilson, and R.W.Mayle, Phys. Rep., {\bf 163}, 63 (1988);
J.Cooperstein, ibid, {\bf 163}, 95 1988).

\bibitem{Bet}
H.A.Bethe,  Rev. Mod. Phys., {\bf 62}, 801 (1990).

\bibitem{Raf}
G.G.Raffelt, {\sl Stars as Laboratories for Fundamental Physics}
(University of Chicago Press, Chicago, 1996).

\bibitem{Taj}
T.Tajima, and K.Shibata, {\sl Plasma Astrophysics}
(Addison-Wesley, Reading MA, 1997).

\bibitem{Pons}
J.A.Pons, et al., ApJ {\bf 513}, 780 (1999).

\bibitem{Ada}
J.B.Adams, M.A.Ruderman, and C.H.Woo, Phys. Rev. {\bf 129}, 1383
(1963).

\bibitem{Gvo}
A.Gvozdev, et al.,  Phys. Rev. D {\bf 54}, 5674 (1996);
M.Kachelriess, and G.Wunner, Phys. Lett. B {\bf 390}, 263 (1997).

\bibitem{Kad}
B.B.Kadomtsev, and V.I.Karpman,  Usp. Fiz. Nauk {\bf 103}, 193
(1971).

\bibitem{Gar}
C.S.Gardner, J.M.Green, N.D.Kruskal, and R.M.Miura,  Phy. Rev.
Lett. {\bf 19}, 1095 (1967).

\bibitem{Mis}
C.W.Misner,  Phys. Rev. Lett. {\bf 19}, 533 (1967).

\bibitem{chi}
H.Y.Chiu, Phys. Rev. {\bf 123}, 1040 (1961).

\bibitem{wei}
S.Weinberg, {\sl Gravitation and Cosmology} (Wiley, New York,
1972).

\bibitem{wig}
E.Wigner, Phys. Rev. {\bf 40}, 749 (1932).

\bibitem{moy}
J.E.Moyal, Proc. Cambr. Phil. Soc. {\bf 45}, 99 (1949).

\bibitem{tap}
F.D.Tappert,  SIAM Rev. {\bf 13}, 281 (1971).

\bibitem{ltsin}
L.N.Tsintsadze, and N.L.Tsintsadze, {\sl Proceedings of the
International Conference on Superstrong Fields in Plasmas}
Varenna, 1997, Ed. by M.Lontano, (American Institute of Physics,
New York, 1998), 170.
\end{references}
\end{document}